\documentclass[pra,twocolumn,preprintnumbers,amsmath,amssymb,superscriptaddress,showpacs,longbibliography]{revtex4-1}
\usepackage[T1]{fontenc} 
\usepackage[utf8]{inputenc} 

\usepackage{graphicx}
\usepackage{latexsym}
\usepackage{amsmath}
\usepackage{graphics}
\usepackage{amssymb}
\usepackage{layout}
\usepackage{verbatim}
\usepackage{amsfonts,epsfig,dsfont}
\usepackage{soul}
\usepackage[dvipsnames,table]{xcolor} 
\usepackage{hyperref}
\hypersetup{colorlinks, linkcolor={blue!90!black}, citecolor={blue!90!black}, urlcolor={blue!90!black} } 

\newcommand{\beq}{\begin{equation}}
\newcommand{\eeq}{\end{equation}}
\newcommand{\bea}{\begin{eqnarray}}
\newcommand{\eea}{\end{eqnarray}}

\newcommand{\tr}{\hbox{Tr}}

\newcommand{\unit}{\mathbb{I}}

\usepackage{color}

\usepackage{ulem} 

\usepackage[]{newtxtext} 
\usepackage[subscriptcorrection,nosymbolsc,smallerops,bigdelims]{newtxmath} 
\DeclareMathAlphabet{\mathcal}{OMS}{cmsy}{m}{n} 
\DeclareMathAlphabet{\mathbcal}{OMS}{cmsy}{b}{n} 
\usepackage{bm}

\begin{document}

\title{A measure of qubit environment
	entanglement for pure dephasing evolutions
}

\author{Katarzyna Roszak}
\affiliation{Department of Theoretical Physics, Faculty of Fundamental Problems of Technology, Wroc{\l}aw University of Science and Technology,
50-370 Wroc{\l}aw, Poland}
\affiliation{Department of Optics, Palack{\'y} University, 77146  Olomouc, Czech Republic}

\date{\today}

\begin{abstract}
	We propose a qubit-environment entanglement measure which is tailored for  
	evolutions that lead to pure dephasing of the qubit, such as
	are abundant in solid state scenarios. 
	The measure 
	can be calculated directly form the density matrix without minimization of any kind. In fact it does not
	require the knowledge of the full density matrix, and it is enough to know the initial
	qubit state and the states of the environment conditional on qubit pointer states.
	This yields a computational advantage over standard entanglement measures, which 
	becomes large when there are no correlations between environmental components in 
	the conditional environmental states.
	In contrast to all other measures of mixed state
	entanglement,
	the measure has a straightforward physical interpretation directly linking the amount of information about the qubit state which 
	is contained in the environment to the amount of qubit-environmnent entanglement.
	This allows for a direct extension of the pure state interpretation of entanglement generated
	during pure dephasing to mixed states, even though pure-state conclusions about qubit 
	decoherence are not transferable.
\end{abstract}
\maketitle

\section{Introduction}
Entanglement for mixed states is hard to understand on an intuitive level.
Already the notion of pure state entanglement \cite{einstein35,schrodinger35b} 
is very abstract,
but still it translates into the existence of information about the joint system state,
which is not contained in states describing each of the subsystems separately.
Defining mixed separable states as states which 
cannot be created by local operations and classical communication (LOCC) \cite{plenio07},
or more simply, by separable operations from product initial states \cite{nielsen00},
and entangled states as states which are not separable, takes the question of what does it mean
if a state is entangled, to a whole new level. Nevertheless, mixed entangled states can be useful,
and there exist some tasks, which will be achieved better 
with the help of a mixed entangled state than by LOCC alone \cite{mesanes06,brandao07,mesanes08,plenio07}.

In system-environment evolutions which lead to pure dephasing (PD) of the system,
pure state entanglement has a particularly meaningful interpretation.
Entanglement, which in this case is directly linked
to system decoherence, describes the amount information about the 
system state which can be extracted from the environment \cite{zurek03,schlosshauer07,Hornberger_LNP09}.
It has been recently shown for mixed states, that without such information transfer,
PD is not accompanied by entanglement
generation \cite{roszak15a,roszak18}. 

We propose a qubit-environment entanglement (QEE) measure which is applicable only for  
PD evolutions with a pure initial qubit state and any state of the environment, which
can be calculated directly from the joint system-environment state.
It actually only requires the knowledge of the initial qubit state 
and the evolution of the environment conditional on qubit pointer states
(hence diagonalization of matrices half the dimension of the joint system-environment state)
and does not require any sort of minimization.
In the situation when there are no correlations between different components
	of the conditional environmental density matrices (such as spins, bosons, etc.), 
the numerical problem reduces to diagonalization of matrices the dimension of each component
separately, which allows to find the evolution of the measure for extremely large environment.
The measure therefore has strong computational advantages.

The proposed measure fulfills the requirements for an entanglement measure \cite{plenio07} within the bounds of its applicability,
as it allows to unambiguously determine separability, reduces to a known entanglement measure for 
pure states, and behaves appropriately under the set of allowed separable operations.
In contrast to all other measures of mixed state
entanglement,
the measure
has a straightforward physical interpretation, directly linking information about the qubit state which 
is contained in the environment to QEE. 
This means that the pure state interpretation of entanglement for PD evolutions \cite{zurek03,schlosshauer07} can be directly extended to mixed states,
even though the link between entanglement and decoherence cannot \cite{Eisert_PRL02,roszak15a,roszak18}.

The paper is organized as follows. In Sec.~\ref{sec2} we introduce the class of systems under study
and the separability criterion characteristic for this kind of evolutions.
In Sec.~\ref{sec3} we introduce the proposed measure for PD entanglement and study its basic properties.
In Sec.~\ref{sec4} we comment on the physical interpretation of the measure linking information 
transfer and entanglement. In Sec.~\ref{sec5} we show the computational advantages of the measure
over standard mixed state entanglement measures and illustrate this with an example in Sec.~\ref{sec6}.
Sec.~\ref{sec7} concludes the paper.

\section{Pure dephasing evolutions \label{sec2}}
We study systems of a qubit and its environment of dimension $N$
initially in a product state with the qubit
in a pure state, 
\begin{equation}
\label{ini}
|\psi\rangle=a|0\rangle +b|1\rangle,
\end{equation}
and no limitations on the initial state of the environment, $\hat{R}(0)$,
so the whole system state can be written as 
$
\hat{\sigma}(0)=|\psi\rangle\langle\psi |\otimes\hat{R}(0)$.
The interaction between them is limited to a class of Hamiltonians 
which lead to PD of the qubit,
and can always be written as
\begin{equation}
\label{ham}
\hat{H}=|0\rangle\langle 0|\otimes\hat{V}_0+|1\rangle\langle 1|\otimes\hat{V}_1,
\end{equation}
where $\hat{V}_i$, $i=0,1$, are arbitrary Hermitian operators 
acting on the environment.
This Hamiltonian can include a free qubit term as long as it commutes 
with the qubit-environment (QE) interaction,
$
\hat{H}_Q=\varepsilon_0|0\rangle\langle 0|+\varepsilon_1|1\rangle\langle 1|$,
so that qubit states $|0\rangle$ and $|1\rangle$ are pointer states \cite{Zurek_PRD81,Zurek_RMP03}.
It can also include a free environment term, $\hat{H}_E$, on which there are 
no limitations.
Hence the operators $\hat{V}_i$ are given by
\begin{equation}
\hat{V}_i=\varepsilon_i\unit_E+\hat{H}_E+\hat{\tilde{V}}_i,
\end{equation}
where $\unit_E$ is the unit operator in the environmental subspace
and the operators $\hat{\tilde{V}}_i$ describe the effect of the bare QE
interaction on the environment. This interaction term must be of the same form as
the total Hamiltonian (\ref{ham}) with $\hat{\tilde{V}}_i$ instead of $\hat{V}_i$.

For this class of Hamiltonians, the QE evolution operator 
can be written in a particularly simple form,
\begin{equation}
\label{evol}
\hat{U}(t)=|0\rangle\langle 0|\otimes\hat{w}_0+|1\rangle\langle 1|\otimes\hat{w}_1,
\end{equation}
with 
\begin{equation}
\label{wi}
\hat{w}_i(t)=\exp(-\frac{i}{\hbar}\hat{V}_i t). 
\end{equation}
This allows to formally find the QE
density matrix at any time, and for the specified initial state we have
\begin{equation}
\label{sigma}
\hat{\sigma}(t)=\left(
\begin{array}{cc}
|a|^2\hat{R}_{00}(t)&ab^*\hat{R}_{01}(t)\\
a^*b\hat{R}_{10}(t)&|b|^2\hat{R}_{11}(t)
\end{array}
\right),
\end{equation}
with 
\begin{equation}
\label{rij0}
\hat{R}_{ij}(t)=\hat{w}_i(t)\hat{R}(0)\hat{w}_j^{\dagger}(t).
\end{equation}
The density matrix is written in matrix form only in terms of qubit pointer states.

As shown previously \cite{roszak15a,roszak18}, for this class of problems
the iff criterion of separability at time $t$ is
(it is also the criterion for zero-discord \cite{roszak18b,chen19})
\begin{equation}
\label{crit}
\hat{R}_{00}(t)=\hat{R}_{11}(t),
\end{equation}
meaning that the state (\ref{sigma}) is separable iff
the state of the environment conditional on one of the two pointer states of the qubit
is the same as the state of the environment conditional on the other pointer state. Otherwise the qubit is entangled with its environment. Separability by no means excludes qubit decoherence 
which is proportional to $\tr\hat{R}_{01}(t)$ (the only exception is 
the situation when the state of the environment is also initially pure; then
decoherence without entanglement is impossible \cite{zurek03,schlosshauer07}).
There are ample examples for realistic qubits undergoing decoherence
both accompanied by QEE \cite{salamon17,strzalka19} and not accompanied by QEE \cite{Eisert_PRL02,Hilt_PRA09,Helm_PRA09,Helm_PRA10,Pernice_PRA11,LoFranco_PRA12,Liu_SR12}.
There are even more examples of systems which undergo PD, which
have never been classified in terms of their entangling or separable nature
\cite{nakamura02a,Cywinski_PRB09,Biercuk_Nature09,Coish_PRB10,Bylander_NP11,Monz_PRL11,Paladino_RMP14,Szankowski_JPCM17,chen18}.

\section{Qubit-environment entanglement measure \label{sec3}}
In the following, we wish to show that the criterion (\ref{crit})
can be used not only to determine if entanglement is present in the system, but can be the basis for an entanglement measure, since the
amount of QEE is proportional to how different 
the environmental states conditional on qubit pointer states,
$\hat{R}_{ii}(t)$, $i=0,1$, actually are,
as suggested by the results of Ref.~\cite{strzalka19}. To this end we propose the quantity
\begin{equation}
\label{meas}
E[\hat{\sigma}(t)]= 4|a|^2|b|^2\left[1-F\left(\hat{R}_{00}(t),\hat{R}_{11}(t)\right)\right]
\end{equation}
as a measure of entanglement for PD evolutions,
or, more generally, for any state that can be written in the form (\ref{sigma}),
as such states can be sometimes obtained for different classes of Hamiltonians
under specific conditions for the initial state \cite{mazurek14b}.
Here $a$ and $b$ are the coefficients of the initial qubit superposition, while the
function $F\left(\hat{\rho}_{1},\hat{\rho}_{2}\right)=
\left[\tr\sqrt{\sqrt{\rho_1}\rho_2\sqrt{\rho_1}}\right]^2$ denotes the Fidelity,
which varies between zero (for $\rho_1\rho_2=0$) and one (for $\rho_1=\rho_2$),
and quantifies similarity between two density matrices.
The measure (\ref{meas}) yields zero for separable QE states and one
when the conditional states of the environment, $\hat{R}_{00}(t)$
and $\hat{R}_{11}(t)$, have orthogonal supports, while the qubit is initially in an equal
superposition state, $|a|=|b|=1/\sqrt{2}$. 

It is important to note that similarity of the measure (\ref{meas}) to the Bures distance in terms of the formula is superficial.
To be used as an entanglement measure, Bures distance requires minimization over the set of all 
separable states \cite{vedral97,vedral98,vedral02,marian03}, while
the proposed measure is directly evaluated from the final density matrix. 
We compare the distance between
two conditional density matrices of the environment, which only requires diagonalization of matrices of the same dimension as the environment. In fact, the proposed measure could be defined using any measure of
distance between density matrices, but the Fidelity allows it to reduce to linear
entropy for pure states.

Note that the measure becomes particularly straightforward to compute
when the conditional environmental density matrices $\hat{R}_{ii}(t)$ retain a product form 
with respect to different components of the environment (when joint qubit-environment evolution
is not accompanied by correlation buildup within the environment),
since $F\left(\hat{\rho}_{1}^a\otimes\hat{\rho}_{1}^b,\hat{\rho}_{2}^a\otimes\hat{\rho}_{2}^b\right)
=F\left(\hat{\rho}_{1}^a,\hat{\rho}_{2}^a\right)F\left(\hat{\rho}_{1}^b,\hat{\rho}_{2}^b\right)$.
This requires the parts of the Hamiltonian describing the free evolution of the environment 
as well as the interaction Hamiltonian to be the sum of terms describing each part of the environment
(the Hamiltonian describing the free evolution of the qubit always has this property),
so that the full Hamiltonian (\ref{ham}) can be written as
$\hat{H}=\sum_k\hat{H}^k$,
where the index $k$ labels the different environmental components.
We can then decompose the conditional evolution operators of the environment, 
$\hat{w}_i(t)=\bigotimes_k\hat{w}_i^k(t)$.
If there are no correlations (classical or quantum) in the initial state of the environment,
$\hat{R}(0)=\bigotimes_k\hat{R}^k(0)$, then the density matrices of the environment conditional
on the qubit pointer state also retain this form, $\hat{R}_{ii}(t)=\bigotimes_k\hat{R}^k_{ii}(t)$.
In this case
\begin{equation}
\label{fid}
F\left(\hat{R}_{00}(t),\hat{R}_{11}(t)\right)
=\prod_k F\left(\hat{R}^k_{00}(t),\hat{R}^k_{11}(t)\right),
\end{equation} 
and much smaller matrices have to be diagonalized to find the value of the PD entanglement measure.
This feature does not simplify the complexity of calculating any of the other measures of mixed
state entanglement.

\subsection{Pure states \label{sec3a}}
Firstly, let us show that the measure (\ref{meas}) reduces to twice the 
(normalized) linear
entropy of the reduced density matrix of either subsystem
for pure states, which is a pure state entanglement
measure \cite{horodecki09}. 
To this end we assume that the initial state of the environment is pure,
$\hat{R}(0)=|R\rangle\langle R|$, since the purity of the initial qubit state is
already assumed in the model, and unitary evolution does not change purity.
The linear entropy of qubit state at time $t$, $\hat{\rho}(t)=\tr_E\hat{\sigma}(t)$, is then
given by
\begin{equation}
\label{sl}
S_L\left[\hat{\rho}(t)\right]= 1-\tr\hat{\rho}^2(t)=
2|a|^2|b|^2\left(1-\left|\langle R|\hat{w}_1^{\dagger}(t)\hat{w}_0(t)|R\rangle\right|^2\right).
\end{equation}
Pure state Fidelity is given by $F\left(|\psi_1\rangle,|\psi_2\rangle\right)
=\left|\langle\psi_1|\psi_2\rangle\right|^2$,
so, since the conditional evolution of the pure initial state of the environment
is given by $|R_{i}(t)\rangle=\hat{w}_i(t)|R\rangle$
(corresponding to the density matrices $\hat{R}_{ii}(t)=|R_{i}(t)\rangle\langle R_{i}(t)|$),
the Fidelity between the two environmental states is 
$
F\left(\hat{R}_{00}(t),\hat{R}_{11}(t)\right)=\left|\langle R|\hat{w}_1^{\dagger}(t)\hat{w}_0(t)|R\rangle\right|^2$,
reducing the QEE measure (\ref{meas}) to twice the linear entropy,
\begin{equation}
\label{epure}
E[\sigma(t)]=2S_L\left[\hat{\rho}(t)\right].
\end{equation}
Hence the measure (\ref{meas}) reduces to a known measure
of pure bipartite entanglement.

\subsection{Separability}
The measure (\ref{meas}) is equal to zero only in three situations;
when either of the initial qubit occupations is equal to zero, $|a|^2=0$ or $|b|^2=0$,
or when the Fidelity between the conditional environmental states 
is equal to one, $F\left(\hat{R}_{00}(t),\hat{R}_{11}(t)\right)=1$, so 
the two states are the same, $\hat{R}_{00}(t)=\hat{R}_{11}(t)$.
During PD evolution,
a qubit initially in one of its pointer states will never become entangled, since the
QE density matrix retains product form.
For all other initial states,
the separability criterion (\ref{crit}) is an iff condition of separability,
so unit Fidelity unambiguously signifies lack of QEE in the system.
Therefore we have
$
E[\sigma(t)]=0$
iff there is no QEE in state $\sigma(t)$
and the measure (\ref{meas}) unambiguously signifies separability.

\subsection{Maximum value of the measure}
The situation when the QEE measure is equal to one is much less straightforward.
It requires the qubit to be initially in an equal superposition state ($|a|=|b|=1/\sqrt{2}$)
and zero Fidelity between the two conditional environmental states. The zero-Fidelity requirement means that $\hat{R}_{00}(t)$ and $\hat{R}_{11}(t)$
have orthogonal supports ($\hat{R}_{00}(t)\hat{R}_{11}(t)=0$)
or, restating the requirement equivalently, that the eigenstates of $\hat{R}_{00}(t)$
with non-zero eigenvalues
occupy a different subspace than the eigenstates of $\hat{R}_{11}(t)$
with non-zero eigenvalues.
Such orthogonality of environmental states, which can naturally
occur during PD evolution, is important for the 
emergence of objectivity \cite{ollivier04,ollivier05,zurek09,korbicz14,horodecki15} and has been studied in detail in this context \cite{roszak19b},
where it has been called ``strict orthogonality''.
The upper limit of the measure (\ref{meas}) is obtained for exactly the same set of states,
for which it would be obtained using any convex-roof entanglement measure, such as Entanglement
of Formation (EOF) \cite{bennett96a,bennett96}.

To show this, let us first study some properties of the QE density matrix
(\ref{sigma})
when the ``strict orthogonality'' condition
is fulfilled, and the qubit is initailly in an equal superposition state.
It is convenient to express all of the $\hat{R}_{ij}(t)$ matrices 
with the help of the environmental state conditional 
on the qubit being in state $|0\rangle$, $\hat{R}_{00}(t)$,
\begin{subequations}
	\label{rij}
	\begin{eqnarray}
	\label{r11}
	\hat{R}_{11}(t)&=&\hat{w}(t)\hat{R}_{00}(t)\hat{w}^{\dagger}(t),\\
	\label{r01}
	\hat{R}_{01}(t)&=&\hat{R}_{00}(t)\hat{w}^{\dagger}(t),\\
	\label{r10}
	\hat{R}_{10}(t)&=&\hat{w}(t)\hat{R}_{00}(t),
	\end{eqnarray}
\end{subequations}
where $
\label{w}
\hat{w}(t)=\hat{w}_1(t)\hat{w}_0^{\dagger}(t)$.
Furthermore,
we can express the matrix $\hat{R}_{00}(t)$ in its eigenbasis,
\begin{equation}
\label{eigen}
\hat{R}_{00}(t)=\sum_nc_n|n(t)\rangle\langle n(t)|.
\end{equation}
The eigenbasis 
can be time-dependent, while
the eigenvalues $c_n$ are not. They are the 
same eigenvalues present in the initial state of the environment,
$
\hat{R}(0)=\hat{R}_{00}(0)=\sum_nc_n|n(0)\rangle\langle n(0)|$,
since the matrices are obtained from one another via a unitary operation.
Since the same logic applies to the other conditional state of the environment
$\hat{R}_{11}(t)$, it is obvious that for the ``strict orthogonality'' condition to be fulfilled,
the number of eigenstates with non-zero eigenvalues of 
the initial state of the environment cannot exceed half of the whole dimension of 
the environment. This in turn limits the initial purity of the environment
to at least twice the minimum purity allowed for a given environment size 
(the ``strict orthogonality'' condition cannot be fulfilled for states
with lower initial purity) \cite{roszak19b}.

Using eqs.~(\ref{rij}) and the basis states (\ref{eigen}) the density matrix (\ref{sigma}) can be written in the concise form
\begin{equation}
\label{sigman}
\hat{\sigma}(t)=\sum_nc_n|\psi_n(t)\rangle\langle\psi_n(t)|,
\end{equation}
where for $a=\frac{1}{\sqrt{2}}$ and $b=\frac{1}{\sqrt{2}}e^{i\varphi}$ we have
\begin{equation}
\label{psin}
|\psi_n(t)\rangle =\frac{1}{\sqrt{2}}\left(|0\rangle\otimes|n(t)\rangle+e^{i\varphi}|1\rangle\otimes
|n_{\perp}(t)\rangle\right).
\end{equation}
The states $|n(t)\rangle$ are all elements of the eigenbasis of $\hat{R}_{00}(t)$
and hence are mutually orthogonal, the same goes for the states $|n_{\perp}(t)\rangle=\hat{w}(t)|n(t)\rangle$,
which are all elements of the eigenbasis of $\hat{R}_{11}(t)$.

The ``strict orthogonality'' condition implies that each state $|n(t)\rangle$
must be orthogonal to each state $|n_{\perp}(t)\rangle$
(otherwise the density matrices $\hat{R}_{ii}(t)$ would not have orthogonal supports).
Hence if $\hat{R}_{00}(t)\hat{R}_{11}(t)=0$, each state $|\psi_n(t)\rangle$
that enters the decomposition (\ref{sigman})
is a maximally entangled bipartite state of Bell type, since
the condition translates into
\begin{equation}
\label{cond}
\forall_n \langle n(t)|n_{\perp}(t)\rangle=0.
\end{equation}
Obviously, it is not a condition which can be fulfilled for more than instances of time 
during the evolution, since the density matrices $\hat{R}_{00}(t)$ and $\hat{R}_{11}(t)$,
which specify the subsets of states $\{|n(t)\rangle\}$ and 
$\{|n_{\perp}(t)\rangle\}$, have to be obtained from the same initial environmental
density matrix $\hat{R}(0)$ via a unitary evolution.
In Sec.~\ref{sec6} we show exemplary evolutions where the condition is fulfilled
at certain points of time.

We will now use the convex-roof construction \cite{vidal00,plenio07}, which allows to generalize pure state entanglement
measures to mixed states by averaging entanglement over the entanglement of the pure state components 
of a mixed state
and additionally minimizing the result over all pure state decompositions,
\begin{equation}
\label{ef}
E_F(\hat{\sigma})=\min_{\{|\phi_i\rangle\}}\sum_ip_i E(|\phi_i\rangle\langle\phi_i|).
\end{equation}
Note that the states $|\phi_i\rangle$, which enter the different decompositions 
$\hat{\sigma}=\sum_ip_i |\phi_i\rangle\langle\phi_i|$ do not need to constitute a basis.
For clarity, we will assume that the pure state entanglement measure is normalized, so it varies 
between zero and one, where zero signifies separability and one is restricted to maximally entangled
states, $E(|\phi\rangle\langle\phi|)\in[0,1]$. The pure state measure based on linear entropy
used before (\ref{epure}) is an example,
as is normalized von Neumann entropy which is used to define EOF. 

For each state $|\psi_n(t)\rangle$ of eq.~(\ref{psin}), we have $E(|\psi_n(t)\rangle\langle\psi_n(t)|)=1$,
so
obviously, for the decomposition (\ref{sigman}) we have
\begin{equation}
\sum_nc_n E(|\psi_n(t)\rangle\langle\psi_n(t)|)=\sum_nc_n=1.
\end{equation}
This by itself does not answer the question if a state of form (\ref{sigma})
would yield the maximum value of EOF, since statistical mixtures of maximally entangled
states can have as little as zero entanglement \cite{roszak10}. 
In general, the minimization over all pure state decompositions requires numerical analysis
and becomes more involved with growing Hilbert space \cite{audenaert01}.
The special feature here is that
each state (\ref{psin}) occupies a different subspace of the QE
Hilbert space, so the matrix (\ref{sigman}) is block diagonal in these subspaces.
Since the class of density
matrices under study is sparse and block-diagonal in $2\times 2$ subspaces, it is 
straightforward to do the minimization analytically. 

The states $|\psi_n(t)\rangle$ all constitute
elements of the eigenbasis of $\hat{\sigma}(t)$ with non-zero eigenvalues,
every possible state to enter each pure-state decomposition 
has to be a normalized linear combination of the states $|\psi_n(t)\rangle$ \cite{Nielsen_Chuang}.
Hence, every state $|\phi_i\rangle$ can be written as
\begin{eqnarray}
|\phi_i\rangle&=&\sum_n\alpha_n^{i}|\psi_n(t)\rangle\\
\nonumber
&=&
\frac{1}{\sqrt{2}}\left(|0\rangle\otimes\sum_n\alpha_n^{i}|n(t)\rangle
+e^{i\varphi}|1\rangle\otimes
\sum_n\alpha_n^{i}|n_{\perp}(t)\rangle\right).
\end{eqnarray}
The environmental state
$\sum_n\alpha_n^{i}|n(t)\rangle$
is orthogonal to its counterpart $\sum_n\alpha_n^{i}|n_{\perp}(t)\rangle$,
since each state in the subset $\{|n(t)\rangle\}$ is orthogonal to each state in the subset
$\{|n_{\perp}(t)\rangle\}$. This means that it is impossible to find a decomposition
which would not yield $\sum_ip_i E(|\phi_i\rangle\langle\phi_i|)=1$.

Hence, because all possible decompositions of state (\ref{sigman})
with maximally entangled $|\psi_n(t)\rangle$ states
involve only maximally entangled pure states, all convex-roof entanglement measures (\ref{ef})
for the whole state will yield their maximum value after minimization, similarly as the
proposed PD entanglement measure (\ref{meas}).
This means that there exists a well defined upper limit to the measure
which coincides with the upper limit of standard mixed-state entanglement measures.

We have already shown that the proposed QEE measure for PD
evolutions (\ref{meas}) is equal to zero iff the QE state is
separable, is equal to one iff EOF would reach its maximum value, and that it
reduces to a good entanglement measure for pure states. 
We will now investigate its other properties.

\subsection{Invariance under local unitary operations}
Firstly, we show invariance under local unitary operations. To this end we will separately
demonstrate invariance in the qubit subspace and in the subspace of the environment.
A unitary operation on the qubit, $U_Q\equiv U_Q\otimes\unit_E$, is equivalent to rotation
of the qubit pointer basis, 
$
|0\rangle \rightarrow |0'\rangle$,
$|1\rangle\rightarrow |1'\rangle$.
This does not change the form of the QE density matrix (\ref{sigma})
regardless if it is done on the initial state of the qubit or on the full density
matrix at time $t$, as long as the qubit basis states are modified accordingly.
Hence,
the measure (\ref{meas}) remains a QEE measure under the operation
and, furthermore, its value remains unchanged,
$E[U_Q\hat{\sigma}(t)U_Q^{\dagger}]=E[\hat{\sigma}(t)]$.

A unitary operation on the environment, $U_E\equiv \unit_Q\otimes U_E$
transforms the density matrix (\ref{sigma})
into
\begin{equation}
\label{sigmau}
U_E\hat{\sigma}(t)U_E^{\dagger}=\left(
\begin{array}{cc}
|a|^2U_E\hat{R}_{00}(t)U_E^{\dagger}&ab^*U_E\hat{R}_{01}(t)U_E^{\dagger}\\
a^*bU_E\hat{R}_{10}(t)U_E^{\dagger}&|b|^2U_E\hat{R}_{11}(t)U_E^{\dagger}
\end{array}
\right).
\end{equation}
Since it is a basic property of the Fidelity that it does not change under
symmetrically applied unitary operations,
$F\left(U\hat{\rho}_{1}U^{\dagger},U\hat{\rho}_{2}U^{\dagger}\right)
=F\left(\hat{\rho}_{1},\hat{\rho}_{2}\right)$,
the entanglement measure (\ref{meas}) yields the same value for the density matrix
(\ref{sigmau}) as it would for (\ref{sigma}). Hence, unitary transformations on the environment
also do not affect the amount of entanglement it signifies.

\subsection{Monotonicity under local operations}
Further study of the properties of the measure (\ref{meas}) under local operations
requires a limitation on the possible operations, since the function (\ref{meas})
does not exist for density matrices which have a different structure than given in eq.~(\ref{sigma}).
For this structure to be maintained, only unitary operations are allowed on the qubit subspace.
There are no such limitations in the subspace of the environment. We find that
if a non-selective quantum operation is performed on the environment, described by a trace-preserving
completely positive map $\Phi$, the value of the QEE measure cannot increase,
\begin{equation}
\label{ineq}
E\left(\Phi(\hat{\sigma}(t))\right)\le E(\hat{\sigma}(t)).
\end{equation}
This is because the operation only affects environmental operators $\hat{R}_{ij}(t)$
in eq.~(\ref{sigma}), so only the Fidelity in the formula for the QEE
measure (\ref{meas}) changes under the operation. The Fidelity cannot decrease under such operations \cite{jozsa94}, so
$F\left(\Phi(R_{00}(t)),\Phi(R_{11}(t))\right)\ge F\left(R_{00}(t),R_{11}(t)\right))$
and consequently the inequality (\ref{ineq}) holds.

	\section{Interpretation \label{sec4}}
	
	 In case of pure QE states, where the qubit and environment interact in such a way that
	 the qubit state undergoes pure dephasing (the interaction is described by a Hamiltonan
	 that is of the type given by eq.~(\ref{ham})), there exists a
	 one-to-one correlation between QEE, qubit decoherence,
	 and the amount of ``which way'' information about the qubit state that has been transferred
	 into the environment \cite{zurek03,schlosshauer07}.
	 This yields a very straightforward interpretation of what it means when a qubit is entangled
	 with its environment. We will restate the argument here using the language introduced above,
	 and show that the argumentation translates into mixed state QEE
	 in terms of information transfer, but not for decoherence.
	 Furthermore, we will show how the evolution of the PD entanglement measure (\ref{meas})
	 describes the process of information transfer.
	 
	 To this end, let us first study the evolution of the purity of the qubit state
	 which describes its level of decoherence.
	 It is given by
	 \begin{equation}
	 \label{purity}
	 \tr\hat{\rho}^2(t)=1-2|a|^2|b|^2\left(1-|\tr\hat{R}_{01}(t)|^2\right),
	 \end{equation}
	 where $\hat{\rho}(t)=\tr_E\hat{\sigma}(t)$ is the density matrix of the qubit obtained
	 after tracing out the degrees of freedom of the environment from the QE density matrix (\ref{sigma}).
	 Initially the qubit state is pure, so its purity is equal to one, but during the evolution
	 it may reach its minimum value, which is equal to $1-2|a|^2|b|^2$, and is obtained when 
	 the off-diagonal elements of the qubit density matrix are equal to zero. These elements 
	 are proportional to $|\tr\hat{R}_{01}(t)|$, so the minimum purity possible for a given 
	 initial state of the qubit is obtained for $|\tr\hat{R}_{01}(t)|=0$.
	 This minimum value depends on the initial qubit state, because this is pure dephasing
	 (so the completely mixed qubit state with $\hat{\rho}(t)=1/2$ can only be obtained for an equal
	 superpostion state). Hence the quantity of interest,
	 which allows to compare the level of coherence between two qubit states 
	 must nontrivially depend on their initial state, since if the qubit is initially in one of the
	 pointer states, it will remain pure throughout the evolution, while an initial
	 equal superposition state may become completely mixed.
	 
	 For pure initial states of the environment	 
	 described by the state $|R\rangle$ as in Sec.~\ref{sec3a} (so that the whole initial QE state is pure), we have 
	 \begin{eqnarray}
	 \nonumber
	 |\tr\hat{R}_{01}(t)|^2&=&\tr\left(\hat{w}_0(t)|R\rangle\langle R|\hat{w}_1^{\dagger}(t)\right)
	 \tr\left(\hat{w}_1(t)|R\rangle\langle R|\hat{w}_0^{\dagger}(t)\right)\\
	 \nonumber
	 &=&\langle R|\hat{w}_1^{\dagger}(t)\hat{w}_0(t)
	 |R\rangle\langle R|\hat{w}_0^{\dagger}(t)\hat{w}_1(t)
	 |R\rangle\\
	 \nonumber
	 &=&\langle R|\hat{w}_1^{\dagger}(t)\hat{R}_{00}(t)\hat{w}_1(t)
	 |R\rangle
	 =\langle R_1(t)|\hat{R}_{00}(t)
	 |R_1(t)\rangle\\
	 \label{trans}
	 &=&
	 F\left(\hat{R}_{00}(t),\hat{R}_{11}(t)\right)
	 \end{eqnarray}
	 since $\hat{R}(0)=|R\rangle\langle R|$ and $\hat{R}_{11}(t)=|R_1(t)\rangle\langle R_1(t)|
	 =\hat{w}_1(t)
	 |R\rangle\langle R|\hat{w}_1^{\dagger}(t)$.
	 Hence for pure QE states, qubit purity depends directly on how different the conditional
	 states of the environment are, which is measured by the Fidelity between them. 
	 
	 This means that
	 there can be no decoherence if no information about the qubit state is present in the environment.
	 The level of pure dephasing (quantified by the reduction of the off-diagonal elements
	 of the qubit density matrix written in the basis of pointer states) depends solely on 
	 the degree of this information transfer (since it is proportional to $|\tr\hat{R}_{01}(t)|$), 
	 but the decoherence in general, which is quantified by the loss of purity, also depends on
	 the initial state of the qubit, which quantifies how much information which can be transferred
	 into the environment is initially present in the state.
	 Pure state QE entanglement is inversely proportional to purity and as such,
	 the generation of entanglement during pure dephasing can be interpreted
	 as a process during which the information about the qubit state is encoded in the environment.
	 
	 For mixed states there is no inherent link between decoherence and the transfer of information
	 about the qubit state into the environment, since the transformation between the conditional
	 density matrices of the environment $\hat{R}_{00}(t)$ and $\hat{R}_{11}(t)$, and 
	 the environmental matrices 
	 responsible for decoherence $\hat{R}_{01}(t)$ as shown for pure states by eqs (\ref{trans}), cannot be made. In fact, pure dephasing
	 can occur as a result of the buildup of classical QE correlations while $\hat{R}_{00}(t)=\hat{R}_{11}(t)$ \cite{Eisert_PRL02,Hilt_PRA09,Helm_PRA09,Helm_PRA10,Pernice_PRA11,LoFranco_PRA12,Liu_SR12,roszak15a}.
	
	On the other hand, the form of the PD entanglement measure (\ref{meas}) directly links the amount
	of QEE present in the system with the Fidelity between the conditional environmental states,
	which quantifies how much information about the system state was transferred into the environment,
	normalized by the amount of information present in the initial qubit state to be transferred
	(as described by the coefficients of the qubit superposition written in the basis of its
	pointer states). This means that although the link between qubit decoherence and QEE generation
	cannot be generalized from pure states to mixed states, the correlation between QEE and information
	transfer can. In fact, the decoherence which is a result of the buildup of classical correlations
	can be distinguished from decoherence resulting from QEE by the study of information transfer
	\cite{roszak15b,roszak18} and this information transfer is experimentally detectable
	in more involved procedures performed on the qubit alone \cite{roszak19a,rzepkowski20}.
	Hence in the case of any PD evolution, QEE generation can be interpreted as a process in
	which information about the qubit state is being encoded in the joint QE state.

	\section{Computational advantage \label{sec5}}
	
	Using the measure (\ref{meas}) to quantify qubit-environment entanglement always 
	requires diagonalization of matrices half the size of those which have to be
	diagonalized to find Negativity of the same state. This is because only the Fidelity between
	conditional density matrices of the environment (\ref{cond}) of dimension $N$ has to be found,
	while finding Negativity requires diagonalization of a matrix of the same size as the whole
	QE system, which is obviously $2N$.
	
	There are situations when the computational advantage of the PD measure becomes much more 
	pronounced due to the properties of the Fidelity. For an environment 
	consisting $K$ qudits, each of dimension $d_k$, where $k$ labels the qudits, 
	the evolution of entanglement will become particularly simple to study 
	if there are no inter-qudit correlations in the conditional states of the environment.
	The lack of correlations at time $t$ means that 
	each conditional state of the environment is of product form with respect to 
	the different qudits,
	\begin{equation}
	\label{prod}
	\hat{R}_{ii}(t)=\bigotimes_{k=1}^K\hat{R}_{ii}^k(t).
	\end{equation}
	The Fidelity can then be obtained using eq.~(\ref{fid}) by calculating the Fidelity of each qudit separately 
	and the value of the PD entanglement measure is then given by
	\begin{equation}
	\label{measprod}
	E[\hat{\sigma}(t)]= 4|a|^2|b|^2\left[1-\prod_{k=1}^{K}F\left(\hat{R}_{00}^k(t),\hat{R}_{11}^k(t)\right)\right].
	\end{equation}
	Finding this value now requires $K$ diagonalizations, but of matrices which are of dimensions
	given by $d_k$, which is always substantially smaller than the whole dimension of the
	environment, $N=\prod_kd_k$.
	Note that this does not particularly simplify the form of the whole QE state at time $t$
	(\ref{sigma}) and is of little help when such entanglement measures as Negativity
	are evaluated, because the product form of the matrices $\hat{R}_{ii}(t)$
	does not translate into a form of the full density matrix which is easier to diagonalize
	(after partial transposition or not).
	
	The obvious scenario which leads to a product form of the conditional environmental density
	matrices (\ref{prod}) is the one where there are no initial inter-qudit correlations,
	\begin{equation}
	\label{iniprod}
	\hat{R}(0)=\bigotimes_{k=1}^K\hat{R}^k(0)
	\end{equation}
	and the full QE Hamiltonian \eqref{ham}
	can be written as a sum over environmental qudits,
	\begin{equation}
	\label{hk}
	\hat{H}=\sum_{k=1}^K\hat{H}_k.
	\end{equation}
	The operators $\hat{V}_i$ in the PD Hamiltonian
	must obviously also have this property,
	\begin{equation}
	\hat{V}_i=\sum_{k=1}^K\hat{V}_i^k,
	\end{equation}
	so the resulting conditional evolution operators of the environment have to be of
	product form
	\begin{equation}
	\hat{w}_i(t)=\bigotimes_{k=1}^K\hat{w}_i^k(t).
	\end{equation}
	Consequently all of the environmental matrices which enter
	the full QE density matrix (\ref{sigma}) have product form
	\begin{equation}
	\label{rijprod}
	\hat{R}_{ij}(t)=\bigotimes_{k=1}^K\hat{R}_{ij}^k(t),
	\end{equation}
	with $\hat{R}_{ij}^k(t)=\hat{w}_i^k(t)\hat{R}^k(0)\hat{w}_i^{k\dagger}(t)$,
	including the conditional density matrices of the environment.
	
	The assumptions of lack of interactions and correlations between constituents of the environment
	are reasonable for many qubits which undergo pure dephasing as their dominating
	decoherence mechanism. For example, for a qubit defined on an NV center interacting
	with an environment of carbon nuclear spins,
	the decoherence occurs on much shorter timescales than nuclear dynamics \cite{Zhao_PRB12,Kwiatkowski_PRB18}. The system is therefore effectively 
	described by a Hamiltonian of the form (\ref{ham}) with an environment of
	noninteracting spin qubits
	(spinful carbon isotopes have spin-$1/2$ nuclei).
	It is straightforward to find systems for which the environment consists of different nulcear spins,
	such as qubits defined on the spin of an electron confined in a quantum dot, where the
	material makeup of the dot stipulates the nuclear spin environment \cite{hanson07,ramsay10} and
	for high magnetic fields such qubits undergo pure dephasing \cite{mazurek14,merkulov02,barnes2011}. 
	Incidentally, for these types of interactions
	to be entangling, the initial state of the environment has to be polarized \cite{strzalka19}.
	An example where the environment is of a different nature consists of an excitonic qubit interacting with a phonon environment
	\cite{borri01,vagov03,vagov04,roszak06b}, which
	is a physical realization of the spin-boson model. Here the environmental qudits (each composed
	of a different phonon mode) are in principle 
	of infinite dimension, 
	but for temperatures up to a few Kelvin, the evolution of each 
	phonon mode can be reliably described using less than ten lowest energy states
	\cite{salamon17}.

	\section{Exemplary evolutions: an environment of non-interacting qubits \label{sec6}}

Let us study a PD evolution between a qubit and an environment consisting $K$ 
qubits,
so $d_k=2$ for all $k$ and the dimension of the whole environment is $N=2^K$. 
The evolution of entanglement will be particularly simple to study using the PD
entanglement measure (\ref{meas}) if there are no correlations between the qubits of the
environment initially and no interaction between them, as described in the previous
section. Such a scenario is 
complex enough to capture the relevant features of QEE, while allowing for its analytical 
description.

We therefore assume that the environmental qubits are initially
in a product state \eqref{iniprod},
and the system qubit interacts with each environmental qubit separately,
so that the full QE Hamiltonian is of the form \eqref{hk}.
To further simplify the analysis of the possible evolutions of QEE
we will assume that the interaction is fully asymmetric, meaning 
that for all $k$, the evolution operators of the environment conditional
on the qubit being in state $|0\rangle$
are equal to unity, $\hat{w}_0^k(t)=\unit$.
This is in fact an assumption that does not limit the generality of the study,
since it is always possible to asymmetrize any PD evolution
by local unitary operations \cite{roszak15a}, and local unitary operations
do not change the amount of entanglement \cite{mintert05,plenio07,horodecki09,aolita15}.

In the following we will always assume that the initial state of the system qubit is an equal 
superposition of its pointer states $|0\rangle$ and $|1\rangle$.

\begin{figure}[tb]
	\includegraphics[width=1\columnwidth]{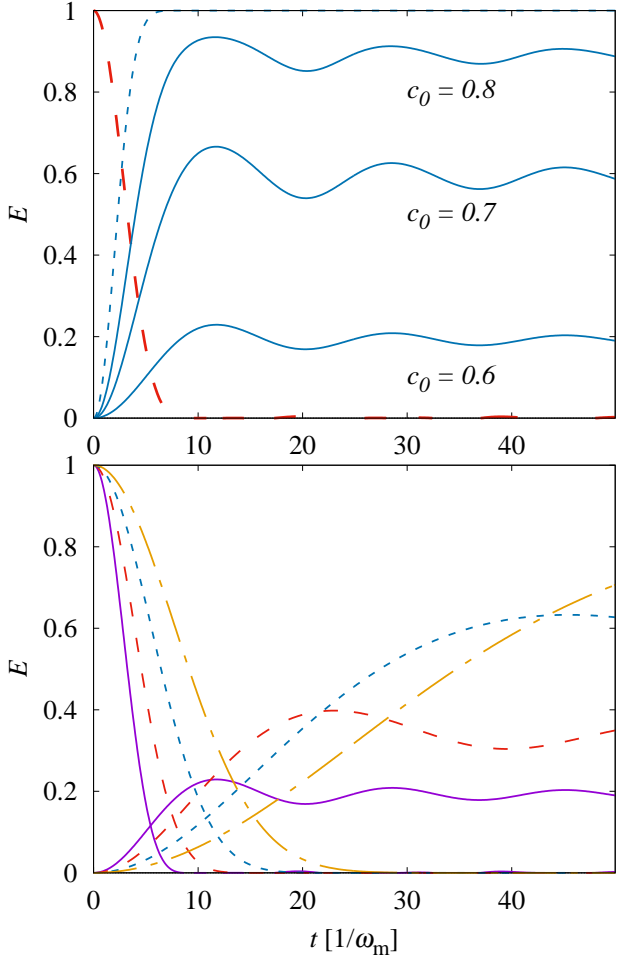}
	\caption{Evolution of QEE and coherence for all environmental qubits in the same initial state.
		Top panel: QEE for $K=10$ environmental qubits at different initial purities
		(solid, blue lines; dotted blue line corresponds to $c_0=1$). Dashed red line shows the
		corresponding coherence of the system qubit. 
		Bottom panel: QEE (curves starting at zero)
		and cohrerence (curves starting at one) for
		a varying number of environmental qubits $K$ for $c_0=0.6$. Solid, purple lines - $K=10$;
		dashed, red lines - $K=20$; dotted, blue lines - $K=40$; dashed-dotted, yellow lines - $K=80$.
	}\label{fig1}
\end{figure}

We choose a simple interaction which would periodically lead to the emergence of a maximally
entangled state in the case of an environment consisting of a single qubit initially in state $|0\rangle$
or $|1\rangle$. To this end, we will consider conditional environmental evolution operators of the form
\begin{equation}
\label{w1k}
\hat{w}_1^k(t)=e^{i\omega_k t}|+\rangle_{kk}\!\langle +|
+e^{-i\omega_k t}|-\rangle_{kk}\!\langle -|,
\end{equation}
with the states $|\pm\rangle_{k}=\frac{1}{\sqrt{2}}\left(|0\rangle_{k}\pm|1\rangle_{k}\right)$
given in the subspace of evironmental qubit $k$.
The basis $\{|0\rangle_{k},|1\rangle_{k}\}$ is the basis in which the initial state of qubit $k$
is diagonal,
$\hat{R}^k(0)=c_0^k|0\rangle_{kk}\!\langle 0|+c_1^k|1\rangle_{kk}\!\langle 1|$.
Here, we have assumed that the interaction of the qubit with each environmental qubit is of the same type,
but the values of $\omega_k$ may differ with $k$.

The coherence of the system qubit in this case is proportional to
\begin{equation}
\label{dekoh}
\tr\hat{R}_{01}(t)=\prod_{k=1}^N\tr \hat{R}_{01}^k(t)=\prod_{k=1}^N\cos\omega_kt
\end{equation}
and is independent of the parameters of the initial environmental qubit states
$c_0^k$ and $c_1^k$. This is a consequence of the choice of interaction together with 
a limitation on the possible initial states of the environment, but it will be 
very convenient, when comparing different scenarios which lead to the same decoherence curves
while the time-evolution of QEE will differ considerably.

It is straightforward to find the conditional evolution of each environmental 
qubit and we get
\begin{align}
\label{er11}
&\hat{R}_{11}^k(t)\\
\nonumber
&=\left(
\begin{array}{cc}
c_0^k\cos^2\omega_kt+c_1^k\sin^2\omega_kt&-i(c_0^k-c_1^k)\sin\omega_kt\cos\omega_kt\\
i(c_0^k-c_1^k)\sin\omega_kt\cos\omega_kt&c_0^k\sin^2\omega_kt+c_1^k\cos^2\omega_kt
\end{array}
\right),
\end{align}
while $\hat{R}_{00}^k(t)=\hat{R}^k(0)$ due to the assumed asymmetry.
In this case we can find the Fidelity for each environmental qubit at time $t$,
which is given by
\begin{equation}
\label{fide}
F\left(\hat{R}_{00}^k(t),\hat{R}_{11}^k(t)\right)=\left(\sqrt{\lambda_+^k}+\sqrt{\lambda_-^k}\right)^2,
\end{equation}
with
\begin{equation}
\lambda_{\pm}^k=\frac{\left(c_0^{k2}+c_1^{k2}\right)\cos^2\omega_kt
+2c_0^kc_1^k\sin^2\omega_kt\pm\sqrt{\Delta^k}}{2}
\end{equation}
and
\begin{eqnarray}
\Delta^k&=&\left(c_0^{k2}-c_1^{k2}\right)^2\cos^4\omega_kt\\
\nonumber
&&+4c_0^kc_1^k\left(c_0^k-c_1^k\right)^2\cos^2\omega_kt\sin^2\omega_kt.
\end{eqnarray}

Before we study an environment composed of many qubits, let us note a couple of the properties of
the single qubit Fidelity between conditional environmental states (\ref{fide}).
Firstly, this Fidelity remains equal to one throughout the evolution only for a maximally mixed initial state
of environmental qubit $k$, $c_0^k=c_1^k=1/2$, for which the joint evolution of the system qubit
and the single environment $k$ would have been separable at all times \cite{roszak15a}.
Hence, the presence of such environmental qubits does not contribute to the value of the PD
entanglement measure (\ref{meas}), even though it does contribute to the decoherence of the system
qubit (\ref{dekoh}). On the other hand, zero Fidelity is only possible if the initial state of
environment $k$ is pure ($c_0^k=0$ or $c_1^k=0$) and will occur only at instances of time when
$\cos\omega_kt=0$. This is in concurrence with the requirement for ``strict orthogonality'' 
proven in Ref.~\cite{roszak19b}, that it is only possible for initial environmental states 
which have at least half of the occupations equal to zero after diagonalization. A qubit state with half of its 
occupations equal to zero is pure.

\begin{figure}[tb]
	\includegraphics[width=1\columnwidth]{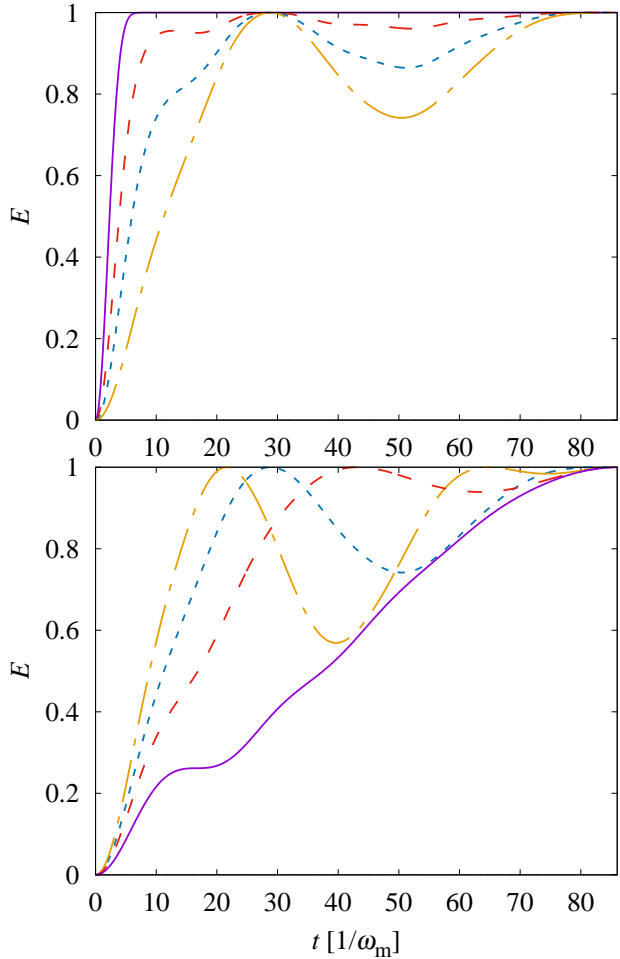}
	\caption{Evolution of QEE for the $k$-th environmental qubit in a pure initial state ($K=10$).
		Top panel: QEE with $k=2$ and 9 qubits in initial state with:
		solid, purple line - $c_0=1$; dashed, red line - $c_0=0.8$; dotted, blue line - $c_0=0.7$; dashed-dotted, yellow line - $c_0=0.6$.
		Bottom panel: QEE for nine environmental qubits in initial state with  $c_0=0.6$
		and qubit $j$ in a pure state with:
		solid, purple line - $j=1$; dashed, red line - $j=2$; dotted, blue line - $j=3$; dashed-dotted, yellow line - $j=4$.
	}\label{fig2}
\end{figure}

Fig.~\ref{fig1} shows the time-evolution of QEE with each environmental qubit initially
in the same state characterized by the occupation $c_0^k=c_0$.
In the upper panel, the number of environmental qubits is set to $K=10$, which means that 
the dimension of the environment $N\eqsim 10^3$. 
The qubits (except for the system qubit) are labeled by the index $k=1,2,\dots, K$
and the constants which govern the conditional evolution of each qubit (\ref{w1k})
were obtained using the formula
\begin{equation}
\label{omegak}
\omega_k=\frac{2\omega_m}{K(K+1)}k,
\end{equation}
so that $\sum_k\omega_k=\omega_m$ regardless of the value of $K$.
The solid, blue lines correspond to mixed states with different values of $c_0\in[1/2,1)$, so that a higher
value corresponds to a greater purity, since the initial purity of the environment
is given by
\begin{equation}
\label{purityenv}
P(\hat{R}(0))=\prod_kP(\hat{R}^k(0))=\left[1-2c_0(1-c_0)\right]^K.
\end{equation}
The dotted, blue line shows the evolution of QEE for a pure
initial state $c_0=1$, while the dashed, red line shows the evolution of coherence given by eq.~\eqref{dekoh}, which is the same for all four QEE evolutions.
Unsurprisingly, the higher the purity of the initial state, the greater values of QEE are reached
during the evolution. Nevertheless, we may observe details of the QEE evolution which are otherwise
unobservable, such as small oscillations, the amplitude of which is damped with higher purity.

The bottom panel of Fig.~\ref{fig1} shows the evolutions of QEE and the corresponding coherence
for a set purity with $c_0=0.6$, but with a varying number of environmental qubits, $K=10,20,40$ and
$80$. $K=80$ means that the environment of dimension $N\eqsim 10^{24}$, yet the results 
are still obtained with little computational effort.

The results of Fig.~\ref{fig2} serve to illustrate the possibility of obtaining the maximum value
of the PD entanglement measure for mixed initial states of the environment. All curves were obtained
for $K=10$. Now the initial state of the environment is modified by the state of a single qubit (labeled
by $j$), which is initially in a pure state with $c_0^j=1$. In the upper panel, the evolution of QEE
is plotted for $j=2$, for different values of $c_0$, which determines
the initial state of all of the other environmental qubits as before.
There is a significant difference in the evolution (the evolution is plotted 
for a longer time, corresponding to half of the time required for the whole QE system
to reach its initial state), even though this change of the initial state of
the environment has not modified the decoherence. Most significantly, points of time are observed
for which the PD entanglement measure reaches its maximum value, even though the QE state
is mixed.

In the bottom panel, the curves have a set purity, but they correspond to different coefficients
$j=1,2,3,$ and $4$. Since different coefficients correspond to different frequencies \eqref{omegak},
this variation leads to significant changes in the QEE evolution.
Note that the decoherence is always the same as is given in the upper panel of Fig.~\ref{fig1}.

\section{Conclusion \label{sec7}}
To conclude, we have proposed a measure of entanglement between a qubit and
an arbitrarily large environment specially tailored to quantify entanglement generated 
during interactions that lead to dephasing of the qubit
(for pure initial qubit states).
The function ranges between zero and one, and
it is zero iff the qubit and environment are in a separable state.
Furthermore, it reduces to a good entanglement measure for pure states, namely the normalized linear 
entropy and has good properties under unitary local operations and the allowed non-unitary
local operations (otherwise the measure is incalculable, but does not yield false results).
It therefore fulfills more than the basic criteria for an entanglement measure.

The measure has a number of advantages compared to other measures, which can be used
to quantify entanglement when one of the studied systems is arbitrarily large.
It can be calculated directly from the QE density matrix
and requires diagonalization of matrices half the dimension of those which need to 
be diagonalized to find Negativity \cite{vidal02,lee00a,plenio05b}.
This may not seem as much, but can be enough to facilitate the transition from an incalculable to a 
calculable quantity for larger environments.
The calculation of the measure does not in fact require the knowledge of the whole QE
state; it requires the initial state of the qubit and the states of the environment conditional
on qubit pointer states. Because of the properties of the Fidelity, the calculation drastically 
simplifies when the evolution does not induce correlations between different components
of the environment.

Furthermore, the measure has a direct physical interpretation, 
setting it apart from all other
measures of mixed state entanglement. Namely, it relates, 
how much the two conditional states of the environment differ from one another,
so consequently, how much information about the state of the qubit has been transferred
into the state of the environment throughout the evolution.
This information is transferred only when the evolution is entangling, the same as 
in case of pure states \cite{zurek03,schlosshauer07}, but contrary to pure states,
the lack of information transfer does not translate to lack of qubit decoherence.

The author is thankful to dr Łukasz Cywiński for helpful discussions.
This work was supported by project 19-22950Y of the Czech Science Foundation.

\end{document}